\documentclass[superscriptaddress,showpacs,aps,prd,nofootinbib]{revtex4}
\usepackage{graphicx}
\usepackage{dcolumn}
\usepackage{amsmath}
\usepackage{epsfig}
\usepackage{colordvi}
\usepackage{color}
\usepackage{hhline}

\begin{document}

\title{\large \bf \boldmath
Study of the process $e^+e^-\to n\bar{n}$ at the VEPP-2000
$e^+e^-$ collider with the SND detector
} 
\author{M.~N.~Achasov}
\affiliation{Budker Institute of Nuclear Physics, SB RAS, 
Novosibirsk, 630090, Russia}
\affiliation{Novosibirsk State University, Novosibirsk, 630090, Russia}
\author{A.~Yu.~Barnyakov}
\affiliation{Budker Institute of Nuclear Physics, SB RAS, 
Novosibirsk, 630090, Russia}
\affiliation{Novosibirsk State University, Novosibirsk, 630090, Russia}
\author{K.~I.~Beloborodov}
\author{A.~V.~Berdyugin}
\author{D.~E.~Berkaev}
\affiliation{Budker Institute of Nuclear Physics, SB RAS, 
Novosibirsk, 630090, Russia}
\affiliation{Novosibirsk State University, Novosibirsk, 630090, Russia}
\author{A.~G.~Bogdanchikov}
\author{A.~A.~Botov}
\affiliation{Budker Institute of Nuclear Physics, SB RAS, 
Novosibirsk, 630090, Russia}
\author{T.~V.~Dimova}
\author{V.~P.~Druzhinin}
\author{V.~B.~Golubev}
\author{L.~V.~Kardapoltsev}
\affiliation{Budker Institute of Nuclear Physics, SB RAS, 
Novosibirsk, 630090, Russia}
\affiliation{Novosibirsk State University, Novosibirsk, 630090, Russia}
\author{A.~S.~Kasaev}
\affiliation{Budker Institute of Nuclear Physics, SB RAS,
Novosibirsk, 630090, Russia}
\author{A.~G.~Kharlamov}
\affiliation{Budker Institute of Nuclear Physics, SB RAS, 
Novosibirsk, 630090, Russia}
\affiliation{Novosibirsk State University, Novosibirsk, 630090, Russia}
\author{A.~N.~Kirpotin}
\affiliation{Budker Institute of Nuclear Physics, SB RAS,
Novosibirsk, 630090, Russia}
\author{I.~A.~Koop}
\affiliation{Budker Institute of Nuclear Physics, SB RAS,
Novosibirsk, 630090, Russia}
\affiliation{Novosibirsk State University, Novosibirsk, 630090, Russia}
\affiliation{Novosibirsk State Technical University,
 Novosibirsk, 630092, Russia}
\author{A.~A.~Korol}
\affiliation{Budker Institute of Nuclear Physics, SB RAS,
Novosibirsk, 630090, Russia}
\affiliation{Novosibirsk State University, Novosibirsk, 630090, Russia}
\author{S.~V.~Koshuba}
\author{D.~P.~Kovrizhin}
\affiliation{Budker Institute of Nuclear Physics, SB RAS,
Novosibirsk, 630090, Russia}
\affiliation{Novosibirsk State University, Novosibirsk, 630090, Russia}
\author{A.~S.~Kupich}
\affiliation{Budker Institute of Nuclear Physics, SB RAS,
Novosibirsk, 630090, Russia}
\affiliation{Novosibirsk State University, Novosibirsk, 630090, Russia}
\author{K.~A.~Martin}
\affiliation{Budker Institute of Nuclear Physics, SB RAS,
Novosibirsk, 630090, Russia}
\affiliation{Novosibirsk State University, Novosibirsk, 630090, Russia}
\author{A.~E.~Obrazovsky}
\author{E.~V.~Pakhtusova}
\affiliation{Budker Institute of Nuclear Physics, SB RAS,
Novosibirsk, 630090, Russia}
\author{Yu.~A.~Rogovsky}
\affiliation{Budker Institute of Nuclear Physics, SB RAS,
Novosibirsk, 630090, Russia}
\affiliation{Novosibirsk State University, Novosibirsk, 630090, Russia}
\author{A.~I.~Senchenko}
\affiliation{Budker Institute of Nuclear Physics, SB RAS,
Novosibirsk, 630090, Russia}
\author{S.~I.~Serednyakov}
\email[e-mail:]{seredn@inp.nsk.su}
\author{Z.~K.~Silagadze}
\author{Yu.~M.~Shatunov}
\affiliation{Budker Institute of Nuclear Physics, SB RAS,
Novosibirsk, 630090, Russia}
\affiliation{Novosibirsk State University, Novosibirsk, 630090, Russia}
\author{D.~A.~Shtol}
\affiliation{Budker Institute of Nuclear Physics, SB RAS, Novosibirsk, 630090,
Russia}
\author{D.~B.~Shwartz}
\affiliation{Budker Institute of Nuclear Physics, SB RAS,
Novosibirsk, 630090, Russia}
\affiliation{Novosibirsk State University, Novosibirsk, 630090, Russia}
\author{A.~N.~Skrinsky}
\author{I.~K.~Surin}
\affiliation{Budker Institute of Nuclear Physics, SB RAS,
Novosibirsk, 630090, Russia}
\author{Yu.~A.~Tikhonov}
\affiliation{Budker Institute of Nuclear Physics, SB RAS,
Novosibirsk, 630090, Russia}
\affiliation{Novosibirsk State University, Novosibirsk, 630090, Russia}
\author{Yu.~V.~Usov}
\affiliation{Budker Institute of Nuclear Physics, SB RAS,
Novosibirsk, 630090, Russia}
\affiliation{Novosibirsk State University, Novosibirsk, 630090, Russia}
\author{A.~V.~Vasiljev}
\affiliation{Budker Institute of Nuclear Physics, SB RAS,
Novosibirsk, 630090, Russia}
\affiliation{Novosibirsk State University, Novosibirsk, 630090, Russia}

\begin{abstract}
 The process $e^+e^-\to n\bar{n}$ has been studied
 at the VEPP-2000 $e^+e^-$ collider with the SND detector in the
 energy range from threshold up to 2 GeV. As a result of the
 experiment, the $e^+e^-\to n\bar{n}$ cross section and
 effective neutron form factor have been measured.
\end{abstract}


\maketitle

\section{Introduction}
Nucleons (neutron and proton) are the subject of theoretical and
experimental studies for many decades. Their internal structure
can be  described in terms of the electromagnetic form factors, electric $G_E$
and magnetic $G_M$, which are complex functions of the momentum transfer 
squared. To measure the nucleon timelike form factors, the reactions
$e^+e^-\to p\bar{p}$, $n\bar{n}$ and $p\bar{p}\to e^+e^-$  are used. 
The $e^+e^-\to B\bar{B}$ cross section, where $ B$ is a spin-1/2
baryon with the mass $m_B$,  is given by the  following expression:
\begin{eqnarray}
\frac{d\sigma}{d\Omega}(s,\theta)=\frac{\alpha^{2}\beta C}{4s}
\left [ |G_M(s)|^{2}(1+\cos^2\theta)+  \frac{1}{\tau}|G_E(s)|^{2}\sin^2\theta
\right ],
\label{eqB1}
\end{eqnarray}
where $s=4E_b^2$, $E_b$ is the beam energy,  $\beta = \sqrt{1-4m_B^2/s}$,
$C$ is a factor taking into account Coulomb interaction of the final 
baryons [$C = y/(1-e^{-y})$ with 
$ y = {\pi\alpha (1+\beta^2)}/{\beta }$
for protons ~\cite{Coulomb}, 
and $C=1$ for neutrons], $\tau = s/4m_B^2$, and $\theta$ is the baryon
polar angle in the $e^+e^-$ center-of-mass (c.m.) frame.
At the threshold  $|G_{E}| = |G_{M}|$.
The total cross section has the following form:
\begin{equation}
\sigma(s) = \frac{4\pi\alpha^{2}\beta C}{3s}
\left [|G_M(s)|^{2} + \frac{1}{2\tau}|G_E(s)|^{2}\right].
\label{eqB2}
\end{equation}
From the measurement of the total cross section the linear combination
of squared form factors 
\begin{equation}
F(s)^2=\frac{2\tau |G_M(s)|^2+|G_E(s)|^2}{2\tau +1}
\label{eqB3}
\end{equation}
can be determined. The function $F(s)$ is called the effective form factor.
It is  this function that is measured in most of $e^+e^-$ and $p\bar{p}$ 
experiments. The $|G_E/G_M|$ ratio can be extracted 
from the analysis of the measured $\cos\theta$ distribution 
(see Eq.(\ref{eqB1})). 

The proton timelike form factor was studied in many experiments. The most
precise measurement of the $e^+e^-\to p\bar{p}$ cross section in the
energy region of interest was performed in the BABAR experiment~\cite{Babar}. 
For the ratio of the proton timelike form factors $|G_E/G_M|$ there are 
two measurements, BABAR~\cite{Babar} and
PS170~\cite{Lear}, which contradict to  each other. For neutron,
the only measurement of the $e^+e^-\to n\bar{n}$ cross section 
was performed in the FENICE experiment~\cite{Fenice}, and there are no data on
the $|G_E/G_M|$ ratio. 

\begin{figure}
\includegraphics[width=0.47\textwidth]{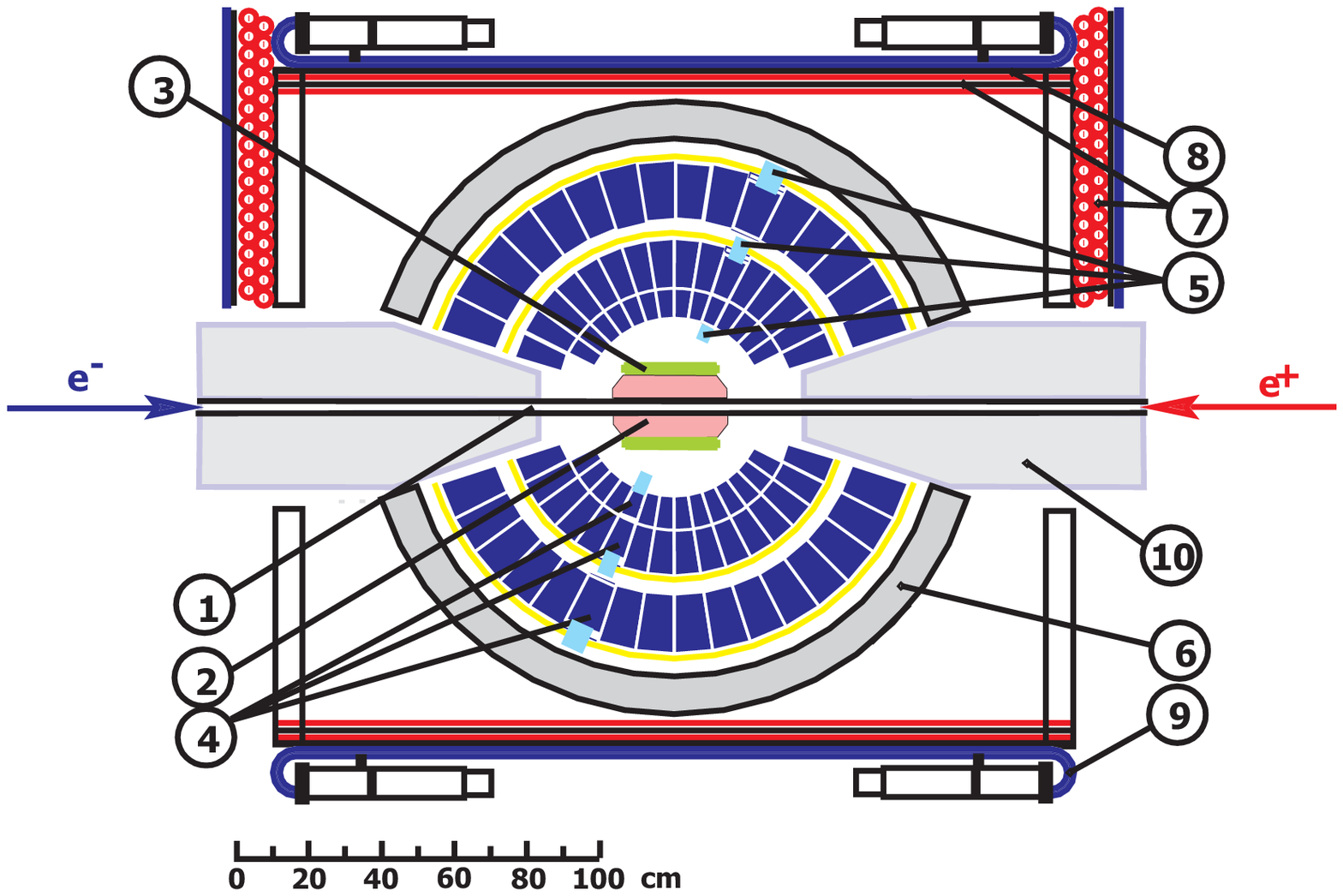} \hfill
\includegraphics[width=0.47\textwidth]{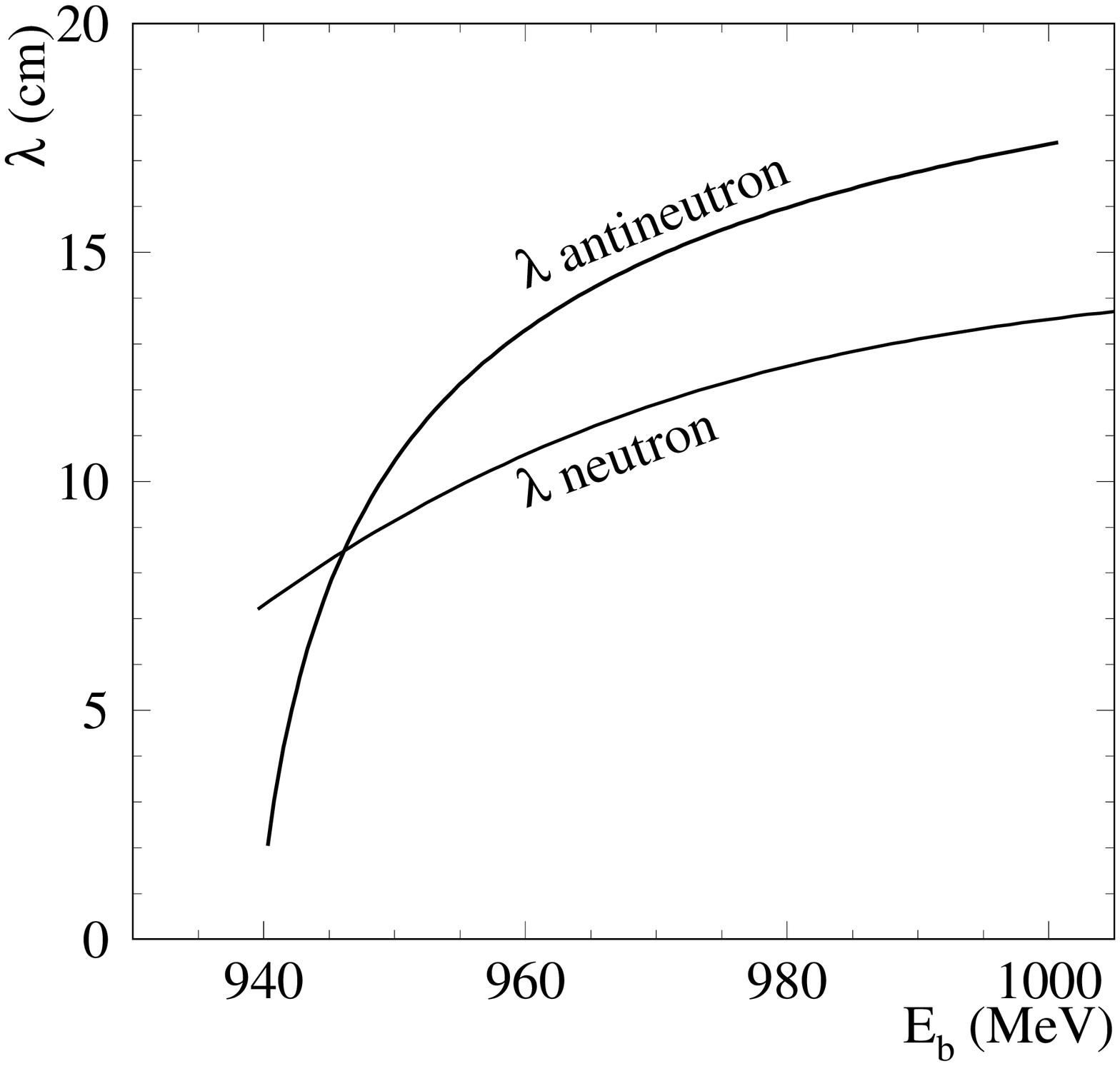} \\
\parbox[h]{0.47\textwidth}{\caption { Schematic view of the SND detector.
The collider vacuum pipe (1) is surrounded by the tracking detector (2) based
on a nine-layer drift chamber. The aerogel Cherenkov counter (3) provides $K$ 
meson identification. The spherical electromagnetic calorimeter consists of 
1680 NaI(Tl) crystals (4) with phototriode (5) readout. The muon detector (7--9)
located after the iron absorber (6) provides muon identification
and suppression of cosmic-ray background.}
\label{Snd2k}} \hfill
\parbox[h]{0.47\textwidth}{\caption {The neutron and antineutron
interaction lengths in NaI(Tl) as a function of the particle energy.}
\label{lann}}
\end{figure}

    In this work we present results on the neutron form factor 
in the c.m. energy range from threshold up to 2 GeV.
The experiment has been carried out at the VEPP-2000  $e^+e^-$ 
collider~\cite{Vepp} with the SND detector~\cite{Snd}
in Novosibirsk. SND (Spherical Neutral  Detector) (Fig.~\ref{Snd2k})
is a general-purpose
nonmagnetic detector for a low energy collider. It
consists of a tracking system, 
a three-layer spherical NaI(Tl) electromagnetic calorimeter (EMC)
and  a muon detector.
Experimental data used in this analysis were accumulated in 2011-2012
in the c.m. energy range 1.8--2.0 GeV.
They correspond to an integrated luminosity of about 10 pb$^{-1}$.
The typical collider luminosity near
the nucleon threshold was about $5\times 10^{30}$ cm$^{-2}$s$^{-1}$.
\begin{figure}
\includegraphics[width=0.47\textwidth]{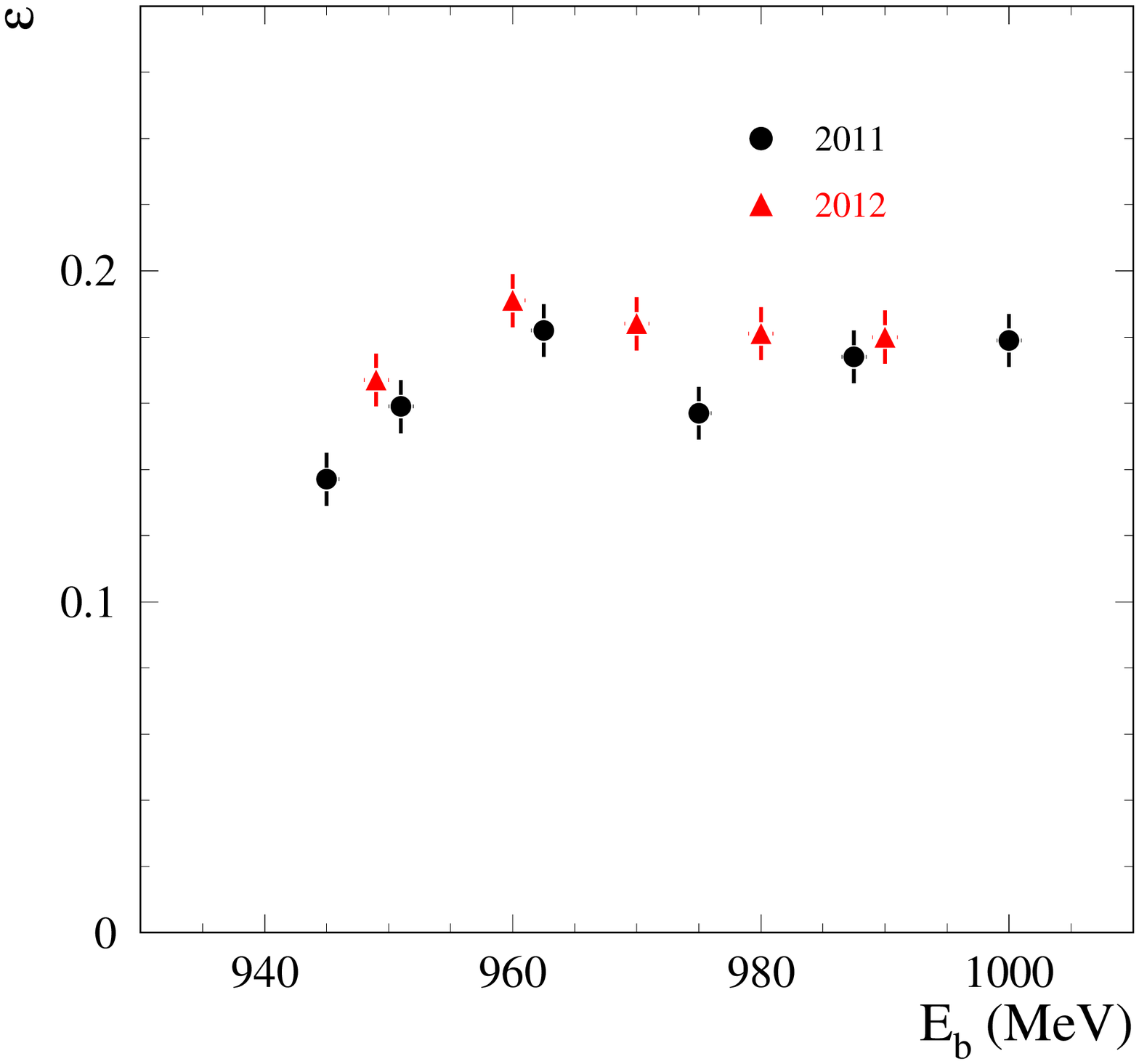} \hfill
\includegraphics[width=0.47\textwidth]{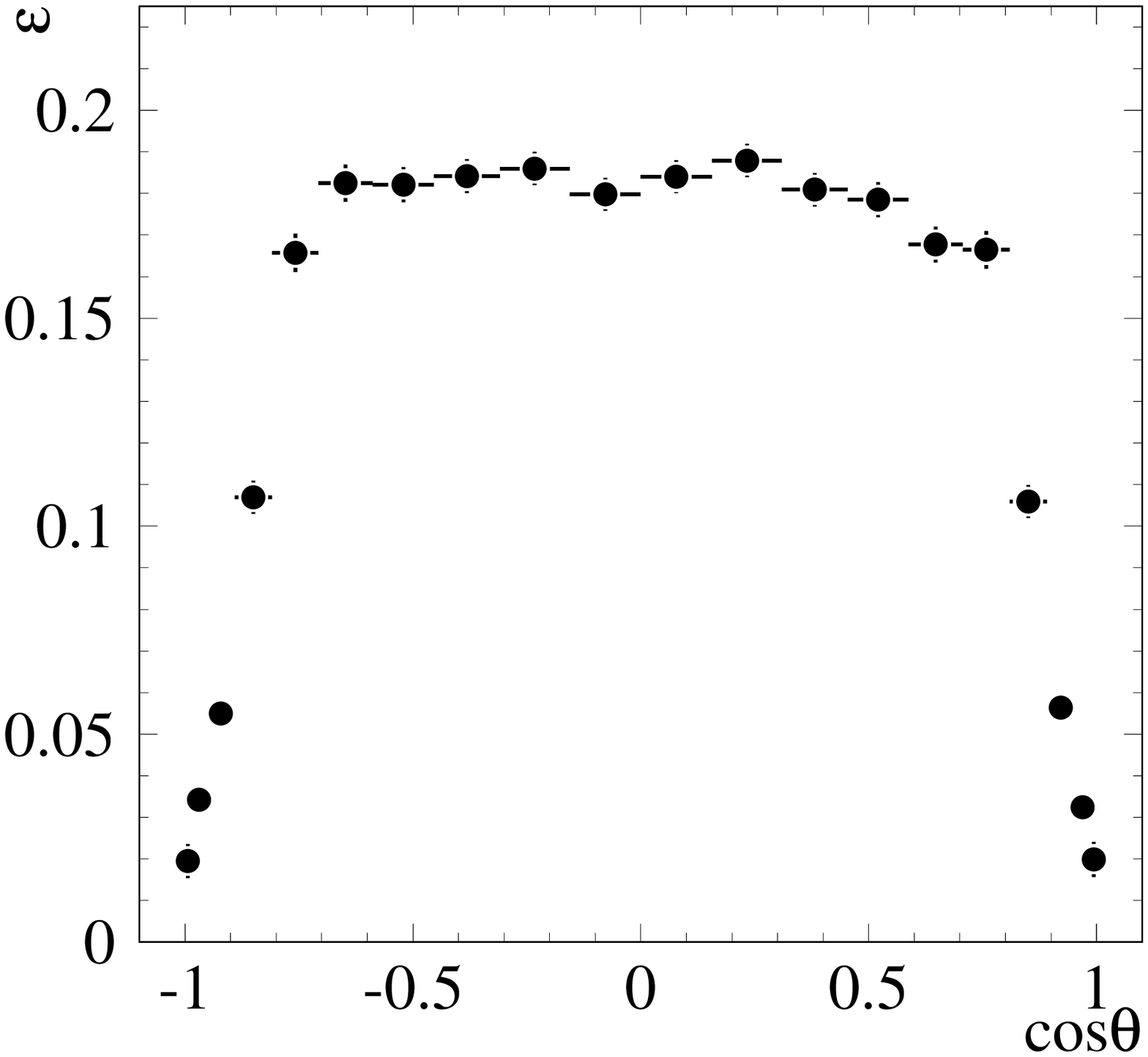} \\
\parbox[h]{0.47\textwidth}{\caption {The energy dependence
of the detection efficiency for $e^+e^-\to n\bar{n}$ events
determined using MC simulation. 
The filled circles show the efficiency for the 2011 data set, 
and the triangles for 2012.}
\label{effmcnn}} \hfill
\parbox[h]{0.47\textwidth}{\caption {The detection efficiency
for $e^+e^-\to n\bar{n}$ events as a function of
$\cos\theta$. The variable $\cos\theta$  bin size is used,
corresponding to $\Delta\theta=9^{\circ}$.}
\label{effcos}}
\end{figure}

\section{Event selection}
The signature of $e^+e^-\to n\bar{n}$ events in the detector is atypical of
$e^+e^-$ annihilation processes. Both final particles, 
neutron and antineutron, cross 
the tracking system  without interaction and give signals deeply inside the 
EMC. So, a $n\bar{n}$ event does not contain
``central'' (originating from the $e^+e^-$ interaction region) charged tracks 
and photons. The neutron interacting in the calorimeter material 
gives a low energy deposition,
while the antineutron annihilates producing several pions with the 
total energy up to  2 GeV. Therefore, the total energy deposition in EMC 
($E_{\rm EMC}$) for
a $n\bar{n}$ event is usually large. But its distribution over the 
calorimeter crystals is strongly nonuniform, i.e. the event momentum  
calculated using energy depositions in the calorimeter crystals ($P_{\rm EMC}$)
significantly differs from zero. The $n\bar{n}$ event looks like several, 
often well separated, clusters (group of adjacent fired  crystals) 
in the EMC. For most events, event-reconstruction
algorithm finds two or more photons. An event may also contain not ``central''
charged track(s).

   In analyses of the $e^+e^-\to n\bar{n}$ process the
value of the antineutron absorption length is of great
significance. The energy dependence of neutron and antineutron
absorption lengths in NaI(Tl) is shown in Fig.~\ref{lann}~\cite{Absorb}.
It is seen that in the VEPP-2000 energy range 
the absorption lengths are much shorter than the effective 
calorimeter thickness, about 40 cm. This leads 
to a high (about 90\% at 2 GeV) absorption efficiency of produced 
particles in the SND detector. 

The selection of $n\bar{n}$ candidates is based on the event properties 
described above. 
We select events with at least two reconstructed photons. An event 
must have a large energy deposition ($E_{\rm EMC}>950$ MeV)
and a large unbalanced momentum in the EMC ($P_{\rm EMC} > 0.5 E_b$).
The condition on $E_{\rm EMC}$ provides full
rejection of beam-background events and significant suppression
of cosmic-ray background.
Most background events from $e^+e^-$ annihilation are 
rejected by the requirement that an event 
may contain only one charged track with $D>0.6$ cm, where $D$ is the distance 
between the charged particle track and the beam axis.

For further reduction of cosmic-ray background we use
the veto from the muon system, the condition that the number of fired EMC
layers in an event equals 3, and the requirement that there is no
cosmic track in the calorimeter. The cosmic track is a group of 
calorimeter crystal hits positioned along 
a straight line with $R_{min}>$10 cm, where $R_{min}$ is a distance
between the track and the detector centre.

To remove the residual background from not correctly reconstructed 
$e^+e^-\to e^+e^-(\gamma),\gamma\gamma (\gamma)$ events 
we require that the fraction of the 
energy deposition in small-angle ($\theta < 36^\circ$ or 
$\theta > 144^\circ$) calorimeter crystals  not exceed 60\%,
and that two most energetic clusters in EMC be not back to back.

The remaining physical background is dominated by the processes
with neutral particles (photons, $\pi^0$'s, neutral kaons)
in the final state, e.g. $e^+e^-\to \gamma\gamma (\gamma)$,  $2\pi^0\gamma$,
$K_SK_L 2\pi^0$.
To suppress the physical background we require
that $E_{\rm EMC}<1500$ MeV, the most energetic  photon in an event has
transverse energy profile not consistent with
the profile expected for the electromagnetic shower~\cite{xinm}, and 
the polar angle of the event momentum 
defined above be in the range $25^\circ<\theta_{P_{\rm EMC}}<155^\circ$.
The latter condition discriminates against multiphoton events
containing extra photons emitted from the initial state at small angles.

   After applying all the selection criteria, the initial number of events,
about $10^9$, recorded  in the energy range 1.8--2.0 GeV is reduced to 
about $5\cdot 10^3$. 

\section{Detection efficiency}
The detection efficiency is determined using Monte-Carlo (MC) simulation.
Its energy dependence is shown in Fig.~\ref{effmcnn} separately for 2011 and 
2012 data sets. At $E_b>960$ MeV 
the efficiency weakly depends on energy and is about 18\% above $E_b=960$ MeV 
and decreases near the $n\bar{n}$ threshold to about 15\%.
The reason for this decrease is
because the annihilation at lower $\bar{n}$ energy occurs
near the center of the detector, and such central events are rejected 
by our selection cuts with a larger probability. 
A nonmonotonic behavior of the detection efficiency as a function
of energy in 2011 and the difference between the efficiencies for 2011 and 
2012 runs are due to variations of experimental conditions during the
data taking period, in particular, due to dead calorimeter channels. 

\begin{figure}
\includegraphics[width=0.47\textwidth]{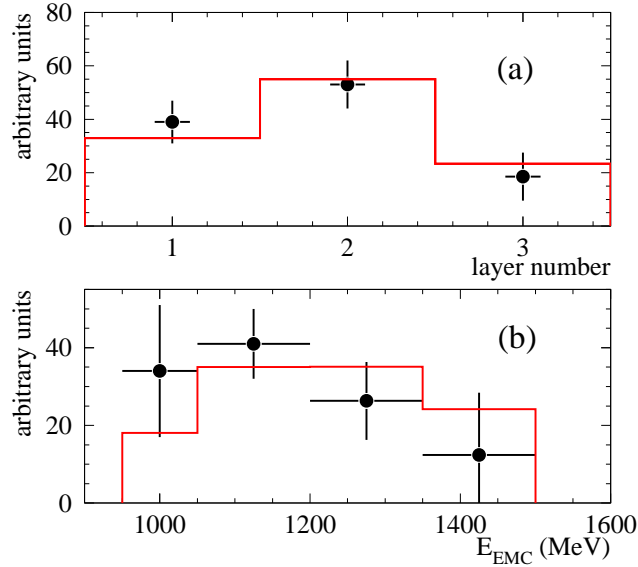}
\caption {(a) The distribution
of the longitudinal position (number of EMC layer) of the crystal
with maximum energy deposition in an $e^+e^-\to n\bar{n}$ event.
(b) The distribution of the total energy deposition in the EMC
for $e^+e^-\to n\bar{n}$ events. The points with error bars represent
data. The histogram is the simulated distribution normalized to the
area of the data distribution.
\label{etemc}}
\end{figure}

The detection efficiency is determined under the assumption that $|G_E|=|G_M|$,
which is true at the threshold. In the BABAR experiment~\cite{Babar} a 
significant deviation of the $|G_E/G_M|$ ratio from unity was observed
in the near-threshold region for the $e^+e^-\to p\bar{p}$ process. 
The ratio reaches 1.5. The deviation from unity is
explained by effects of final state interaction~\cite{Dmitriev}. A similar
deviation is expected for neutron. The model dependence in the detection 
efficiency arises from limited detector acceptance. The detection efficiency 
as a function of $\cos{\theta}$ is shown in  Fig.~\ref{effcos}. 
The efficiency has a plateau in the range  
$36^\circ<\theta<144^{\circ}$, corresponding to $|\cos{\theta}|<0.8$.
The difference (3\%) between the detection 
efficiencies determined with $|G_E/G_M|=1.5$ and $|G_E/G_M|=1$ is taken 
as an estimate of the model uncertainty. 

Not quite perfect simulation of detector response for antineutrons
may lead to systematic shift in the detection efficiency.
In Fig.~\ref{etemc}(a) we compare data and simulation distributions
of the longitudinal position (number of EMC layer) of the crystal 
with maximum energy deposition in an $e^+e^-\to n\bar{n}$ event.
To obtain the data distribution we measure the average over energy
points visible cross section for each of the three bins using the
procedure described in Sec.~\ref{xsec} and subtract physical
background. 
Since the data and simulated distributions are in agreement, we conclude 
that the probability of antineutron absorption in EMC is reproduced by the 
simulation reasonably well.

In Fig.~\ref{etemc}(b) the distribution of the total energy deposition
in the EMC is shown. Although the difference between the data and simulated
distributions is not statistically significant, we interpret it as an 
indication of imperfect simulation.
To reach better agreement, we shift the simulated spectrum to left by 
about 50 MeV. This leads to decrease of the detection efficiency by 10\%.
This value is taken as an estimate of the systematic uncertainty due to
the condition on $E_{\rm EMC}$.

For other selection parameters (the total event momentum, the photon shower
profile, the fraction of the energy deposited at small polar angles, etc.),
we vary cut boundaries over wide ranges and determine variations of the 
measured cross section. The variations summed in quadrature are about 10\%.
A total systematic uncertainty in the detection efficiency including 
the model uncertainty and the uncertainty due to imperfect simulation
of the detector response is estimated to be 14\%.

\section{Angular distribution}
\begin{figure}
\includegraphics[width=0.47\textwidth]{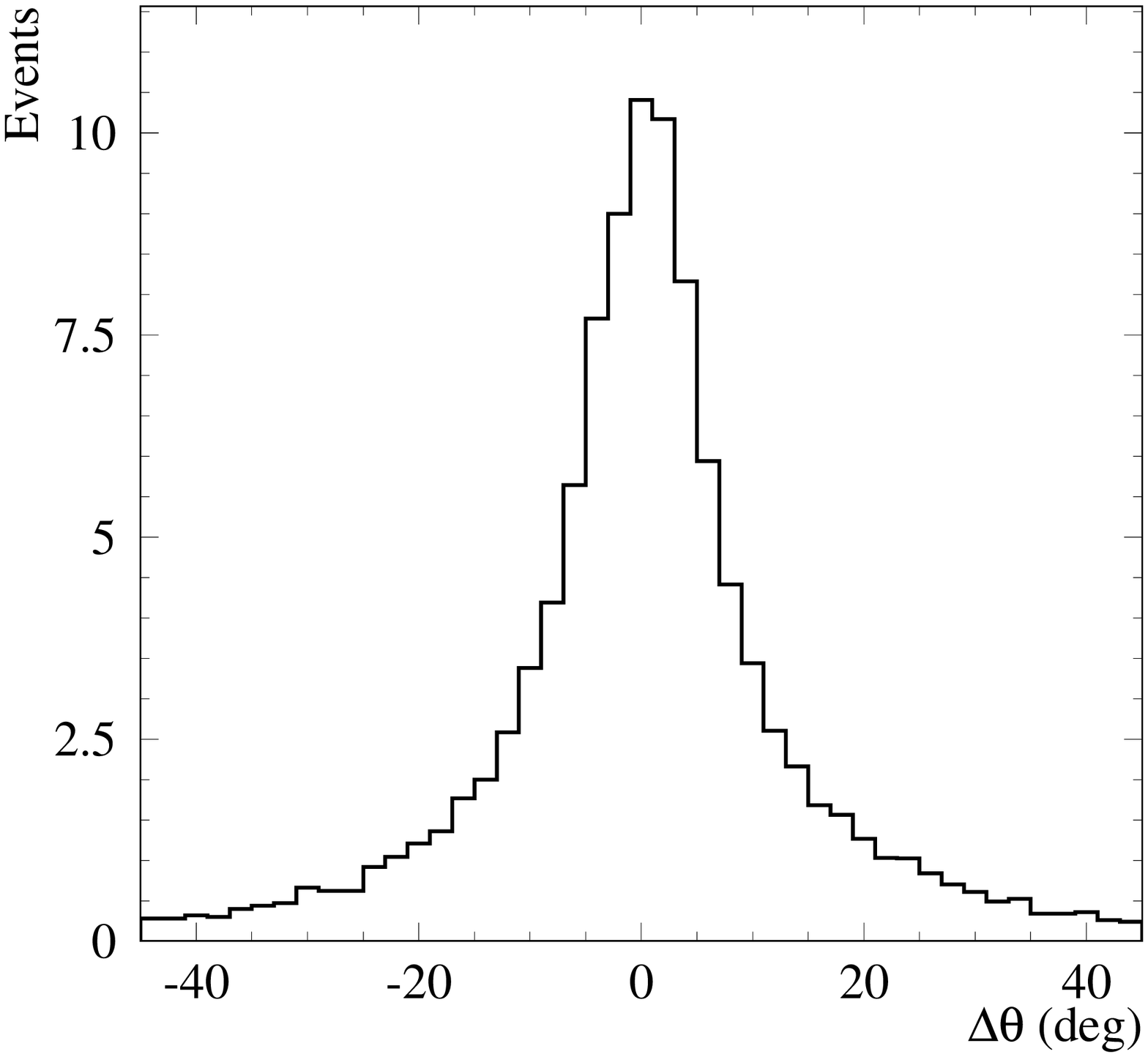} \hfill
\includegraphics[width=0.47\textwidth]{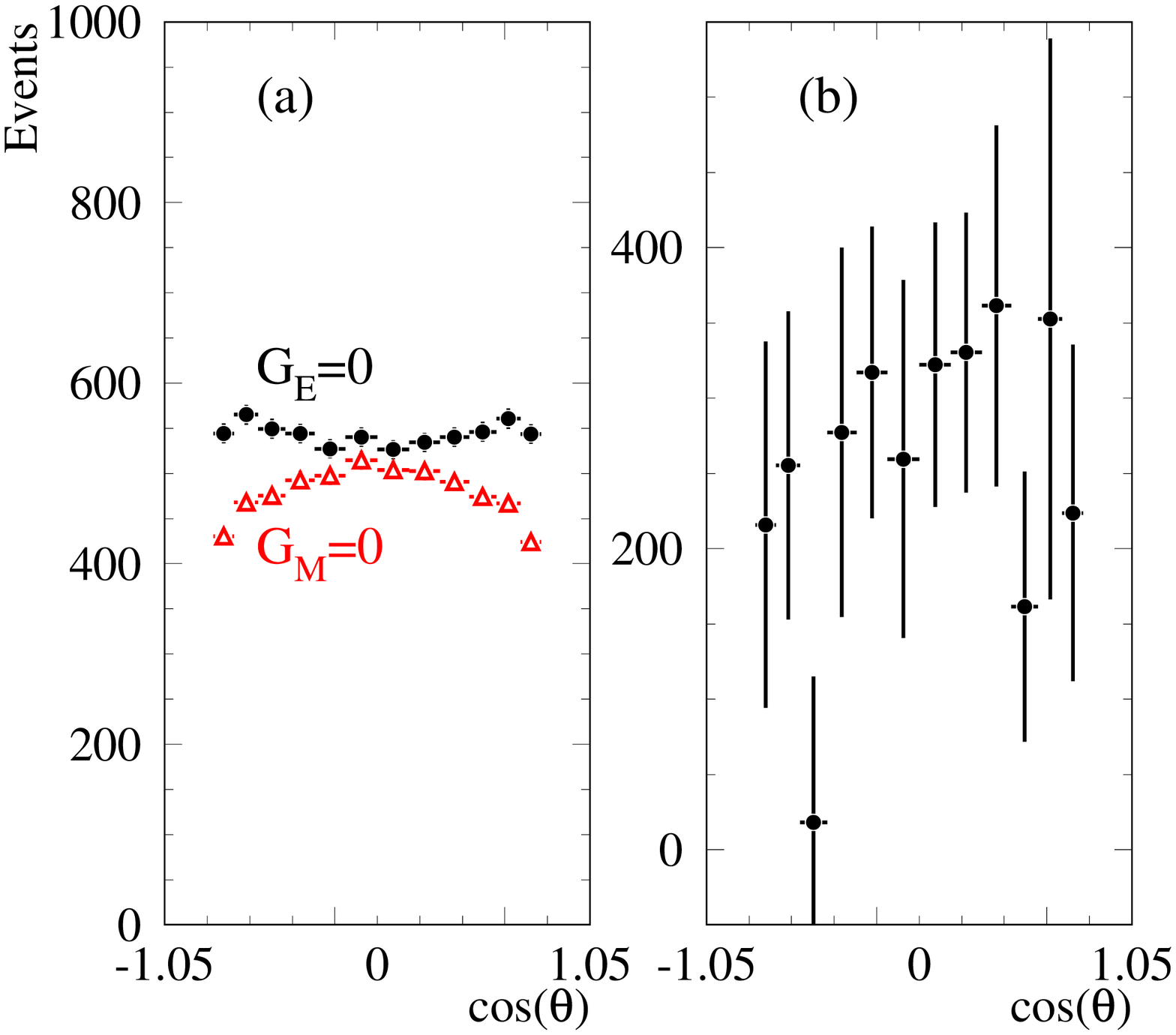} \\
\parbox[h]{0.47\textwidth}{\caption {The distribution of
the difference between the true and measured antineutron
polar angles at $E_{b}=960$ MeV.}
\label{dangl}} \hfill
\parbox[h]{0.47\textwidth}{\caption {(a) The $\cos{\theta}$ distribution 
for simulated $e^+e^-\to n\bar{n}$ events 
generated with $G_E=0$ and $G_M=0$. (b) The $\cos{\theta}$ 
distribution for data $e^+e^-\to n\bar{n}$ events. }
\label{nngegm}}
\end{figure}

The antineutron looks as a wide cluster or several clusters
in the calorimeter. The polar angle of the calorimeter
crystal with maximum energy deposition is used as an
estimate of the antineutron polar angle.
The distribution of the difference between the true and measured 
antineutron polar angles for simulated $n\bar{n}$ events
is shown in Fig.~\ref{dangl}.
The RMS of this distribution is about  8$^{\circ}$.
About 70\% of the reconstructed $n\bar{n}$ events are located
within $\pm 15^{\circ}$ of the true antineutron direction.

The simulated $\cos\theta$ distributions obtained 
using the event samples with $G_M=0$ and $G_E=0$
are shown in Fig.~\ref{nngegm}(a). 
The $\cos\theta$ distribution for data
$n\bar{n}$ events is shown in Fig.\ref{nngegm}(b).
It is seen that the current level of statistics
does not allow us to determine the $|G_E/G_M|$ ratio 
from experiment.

\begin{figure}
\includegraphics[width=0.47\textwidth]{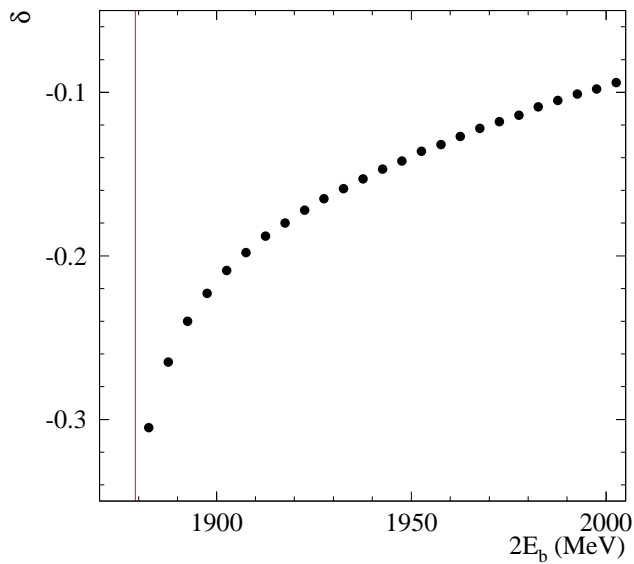} \\
\caption {The energy dependence of the radiative 
correction for the $e^+e^-\to n\bar{n}$ process. The vertical line 
indicates the $n\bar{n}$ threshold.
\label{frad}}
\end{figure}

\section{Cross section\label{xsec}}
The sample of selected $n\bar{n}$ candidates contains a significant fraction,
about 70\%, of cosmic background events. To separate contributions of cosmic
and $e^+e^-$ annihilation events we use a feature of our experiment that data
were collected during about 1200 independent runs with different average 
luminosity ranged from $1\times10^{30}$ 
to $8\times10^{30}$ cm$^{-2}$s$^{-1}$. The
number of selected $n\bar{n}$ candidates in the $i$th run can be written as 
\begin{equation}
N_i = xT_i+\sigma_{vis}(E_b)L_i,
\label{Nev}
\end{equation}
where $x$ is the cosmic background rate,
which is assumed to be constant during the experiment, $T_i$ and $L_i$ are 
the run duration and integrated luminosity, respectively,
and $\sigma_{vis}$ is the visible 
cross section for $e^+e^-$ annihilation events passed our selection,
which is a constant for runs belonging a specific energy point.
The system of equations~(\ref{Nev}) is solved using the maximum-likelihood 
method independently for the 2011 and 2012 experiments. As a result, we
obtain the values of the visible cross section for 7 points below
the $n\bar{n}$ threshold and 11 points above. The values of
cosmic rates in 2011 and 2012 are found to be compatible to each other.
The average $x$ value is found to be $(1.40\pm 0.07)\times 10^{-3}$ Hz.

     The measured values of $\sigma_{vis}$ are used to obtain
the $e^+e^-\to n\bar{n}$ cross section: 
\begin{equation}
\sigma_{n\bar{n}} = 
\frac{\sigma_{vis}-\sigma_{vis,p\bar{p}}-{\sigma}_{vis,bkg }}
{ \varepsilon (1+\delta) },
\label{snn}
\end{equation}
where $\varepsilon$ is the detection efficiency, $\delta$ is a radiative 
correction, $\sigma_{vis,p\bar{p}}$ is the visible cross section for
$e^+e^-\to p\bar{p}$ events passed our selection criteria,
$\sigma_{vis,bkg}$ is the visible cross section for other
background processes. The radiative correction is calculated according to 
Ref.~\cite{radcor} assuming that the $e^+e^-\to n\bar{n}$ cross section is 
a constant in the energy region of interest. The systematic uncertainty 
due to this assumption is estimated to not exceed 0.02.
The energy dependence of the radiative correction is shown in Fig.~\ref{frad}.

The $e^+e^-\to p\bar{p}$ background contribution is calculated as
$\sigma_{vis,p\bar{p}}=\sigma_{p\bar{p}}\varepsilon_{p\bar{p}}\delta_{p\bar{p}}$,
where the Born cross section $\sigma_{p\bar{p}}\approx 0.85$ nb is
taken from Ref.~\cite{Babar}, the radiative correction 
$\delta_{p\bar{p}}\simeq\delta$, and the detection efficiency 
$\varepsilon_{p\bar{p}}\approx 0.01\varepsilon$.
We estimate the systematic uncertainty on the $p\bar{p}$ contribution 
to be about 30\%.

The background contribution from physical processes other than
$e^+e^-\to p\bar{p}$ (${\sigma}_{vis,bkg }$) is measured directly 
below the $n\bar{n}$ threshold.
Its value averaged over 7 energy points ranged from $2E_b=1.8$ to 1.87 GeV 
is found to be $15\pm11$ pb, about 10\% of $\sigma_{vis}$ above
threshold. This value is in agreement with the background estimation 
($10\pm5$ pb) from MC simulation for the processes 
$e^+e^-\to \gamma\gamma (\gamma)$,  
$2\pi^0\gamma$, $3\pi^0\gamma$, $K_SK_L$, $K_SK_L\pi^0$, and $K_SK_L 2\pi^0$.
To obtain the hadronic cross sections we 
use the experimental data from Refs.~\cite{exps} and isotopic relations.
In both MC simulation and data we do not observe strong energy dependence 
of the background cross section. Therefore, the average value of
$\sigma_{vis,bkg }$ determined below threshold is taken as an estimate of 
background above threshold. An additional systematic 
uncertainty of 10 pb is introduced to account for a possible energy 
dependence of the background.

\begin{figure}
\includegraphics[width=0.47\textwidth]{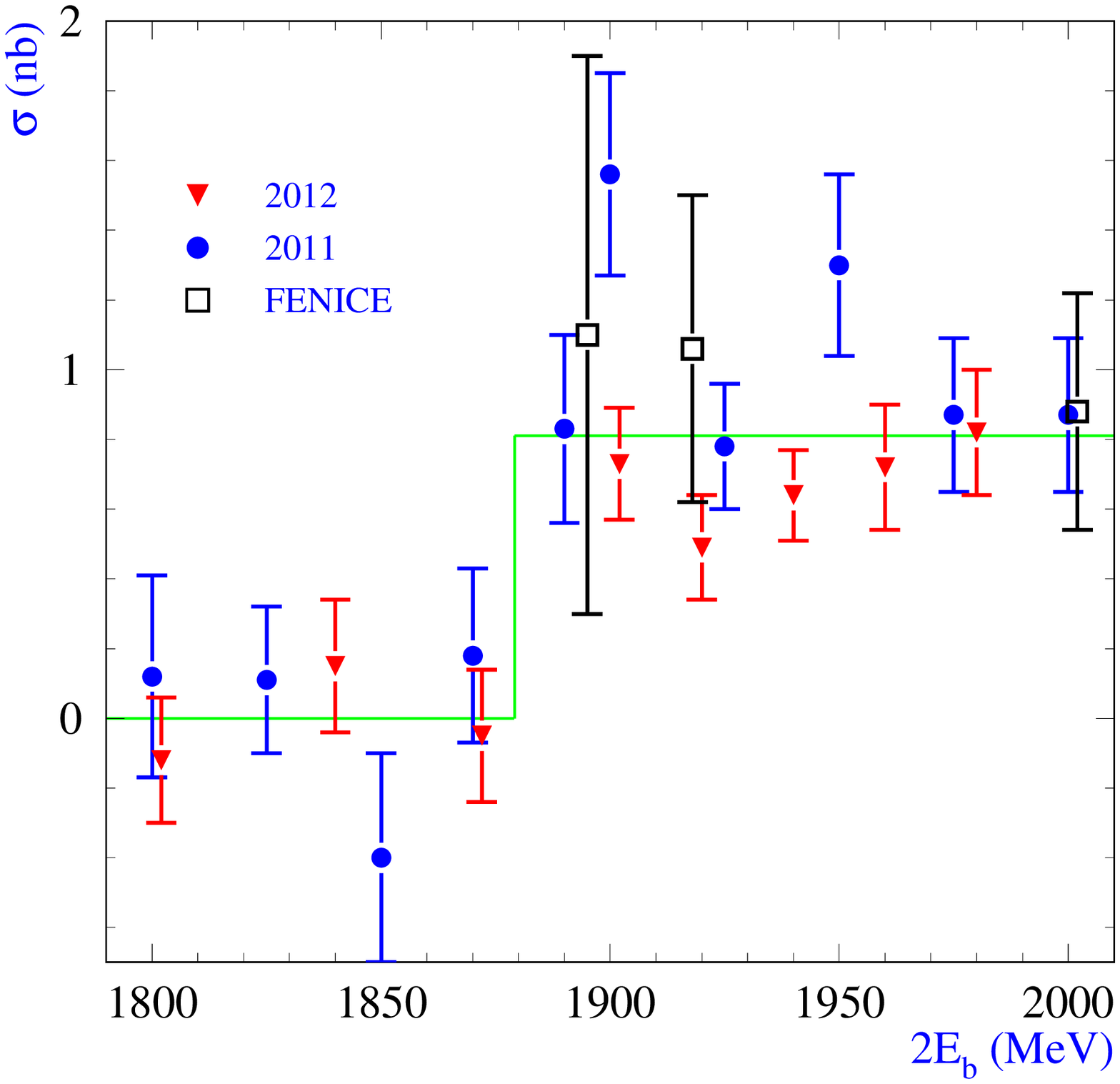} \hfill
\includegraphics[width=0.47\textwidth]{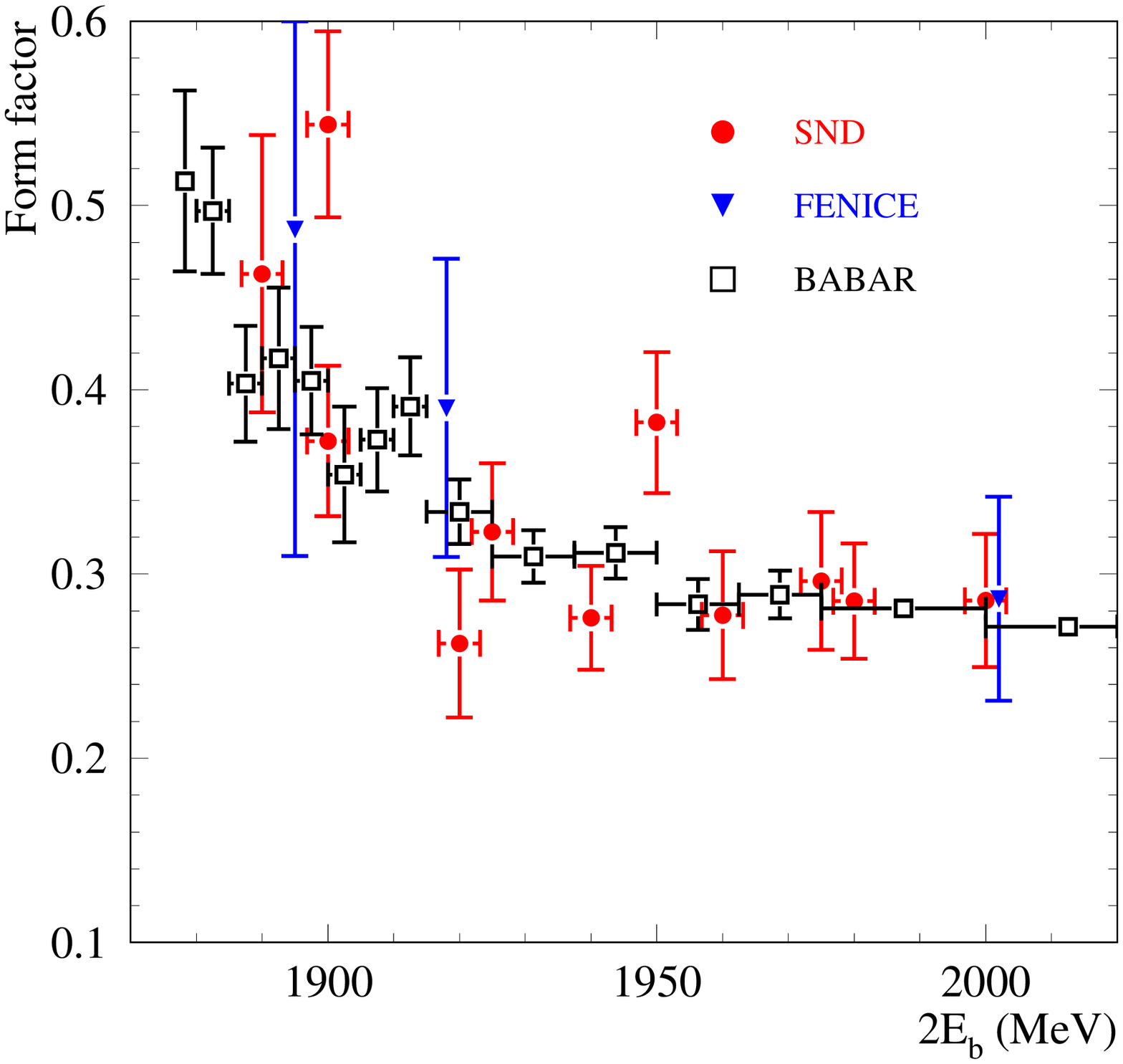} \\
\parbox[h]{0.47\textwidth}{\caption { 
The $e^+e^-\to n\bar{n}$ cross section measured in this work.
The filled triangles represent 2011 data, while the filled circles 
correspond to 2012 data. The FENICE results~\cite{Fenice} are shown by
the empty squares. The lines below and above the $n\bar{n}$ threshold
indicate the average levels of the cross section.
The quoted errors are statistical}.
\label{snnb}} \hfill
\parbox[h]{0.47\textwidth}{\caption {A comparison of the neutron effective
form factor measured in this work (SND) and in Ref.~\cite{Fenice} (FENICE)
and the proton effective form factor measured in the BABAR 
experiment~\cite{Babar}.} 
\label{formnn}}
\end{figure}

   The values of the $e^+e^-\to n\bar{n}$ Born cross section 
obtained using Eq.~(\ref{snn}) are listed in Table \ref{tab-1} and
shown in Fig.\ref{snnb} in comparison with the previous
measurement~\cite{Fenice}. It is seen that our 2011 and 2012 data
and the FENICE results are in reasonable agreement.

    The systematic uncertainty on the measured cross section includes
the uncertainty on the background subtraction (0.05 nb), 
the uncertainty on the detection efficiency (0.12 nb),
the uncertainties in the integrated luminosity (0.02 nb)
and the radiative correction (0.02) nb.
The total systematic error is 0.14 nb or 17\% of the cross section.
The error in the cosmic background subtraction (0.12 nb)
is included into the statistical error $\sim$25\%.
\begin{table*}
\caption{The $e^+e^-\to n\bar{n}$ cross section ($\sigma_{n\bar{n}}$) 
and neutron effective form factor ($F_n$) measured in this work.
The quoted errors are statistical. The systematic error 
is 17\% for the cross section and 9\% for the form factor.
\label{tab-1}}
\begin{ruledtabular}
\begin{tabular}{ccccc}
N & Experiment & $2E_b$, MeV& $\sigma_{n\bar{n}}$, nb& $F_n$\\
\hline
1 & 2011 & 1890 & 0.83$\pm$0.27 & 0.45$\pm$0.09 \\
2 & 2011 & 1900 & 1.56$\pm$0.29 & 0.53$\pm$0.06 \\
3 & 2011 & 1925 & 0.78$\pm$0.18 & 0.32$\pm$0.04 \\
4 & 2011 & 1950 & 1.30$\pm$0.26 & 0.38$\pm$0.04 \\
5 & 2011 & 1975 & 0.87$\pm$0.22 & 0.29$\pm$0.04 \\
6 & 2011 & 2000 & 0.87$\pm$0.22 & 0.28$\pm$0.04 \\
7 & 2012 & 1900 & 0.73$\pm$0.16 & 0.37$\pm$0.06 \\
8 & 2012 & 1920 & 0.49$\pm$0.15 & 0.27$\pm$0.06 \\
9 & 2012 & 1940 & 0.64$\pm$0.13 & 0.28$\pm$0.04 \\
10 & 2012 & 1990 & 0.72$\pm$0.18 & 0.28$\pm$0.05 \\
11 & 2012 & 1980 & 0.82$\pm$0.18 & 0.29$\pm$0.05 \\
\end{tabular}
\end{ruledtabular}
\end{table*}

The measured $e^+e^-\to n\bar{n}$ cross section has unusual 
behavior: it is approximately constant in the energy range from 
threshold up to 2 GeV. 
Similar behavior in the near threshold region was observed for 
the $e^+e^-\to p\bar{p}$ cross section~\cite{Babar}.
The average $e^+e^-\to n\bar{n}$ cross section below 2 GeV, about 0.8 nb, 
is close to the average cross section for $e^+e^-\to p\bar{p}$, 0.85 nb.

From the measured cross section we determine the effective neutron
form factor [Eq.(\ref{eqB3})].
The form-factor energy dependence is shown in Fig.~\ref{formnn}
in comparison with the previous FENICE measurements~\cite{Fenice}, and the
proton form-factor data~\cite{Babar}. Both neutron and proton form factors 
increase near threshold and are close to each other within the 
measurement errors.

\section {Summary} 
In the experiment with the SND detector  at the VEPP-2000 $e^+e^-$ collider 
the $e^+e^-\to n\bar{n}$ cross section and the neutron effective
form factor have been measured in the c.m. energy range 
from the $n\bar{n}$ threshold up to 2 GeV.
The obtained results are in agreement with the previous FENICE 
measurements~\cite{Fenice} but more precise.

 This work is partially supported  in the framework of the State 
order of the Russian Ministry of Science and Education
and by RFBR grants No. 12-02-00065-a, 13-02-00375, 14-02-31375-mol-a 
and scientific school grant 2479.2014.2.

\end{document}